\documentclass{article}
\usepackage[utf8]{inputenc}
\usepackage{graphicx}
\usepackage{times}
\usepackage{xcolor}
\usepackage{amsmath}
\usepackage{subfigure}
\usepackage{amssymb}
\usepackage[numbers,sort&compress]{natbib}
\usepackage{multirow}
\newcommand{\be}{\begin{equation}
\newcommand{\ee}{\end{equation}}}
\newcommand{\bea}{\begin{eqnarray}}
\newcommand{\eea}{\end{eqnarray}}
\newcommand{\nn}{\nonumber}
\begin{document}

\title{Analytic study of conformable  Schrodinger equation with Hydrogen atom}

\author{Mohamed.Al-Masaeed, Eqab.M.Rabei and Ahmed Al-Jamel\\
Physics Department, Faculty of Science, Al al-Bayt University,\\ P.O. Box 130040, Mafraq 25113, Jordan\\moh.almssaeed@gmail.com\\eqabrabei@gmail.com\\
aaljamel@aabu.edu.jo, aaljamel@gmail.com}

\maketitle

%\begin{history}
%\received{Day Month Year}
%\revised{Day Month Year}
%\end{history}

\begin{abstract}
In this paper, the  conformable Schrodinger equation for hydrogen atom with given conformable potential is solved. The conformable wave functions and energy levels are obtained, and the traditional energy levels and wave function for hydrogen atom are recovered when $\alpha = 1$. The probability density for the first three levels and different values of $\alpha$ is plotted. It is observed that the probability density gradually convert from $\alpha=0.5$ to $\alpha = 1$ for all levels.
\\

\end{abstract}

\section{Introduction}
The Schrodinger equation, which is the quantum equivalent of Newton's second law in classical mechanics, is an important conclusion in quantum mechanics for obtaining the wave function. Only a few idealized systems, like the hydrogen atom, may achieve the exact solutions to the Schrodinger equation. And the method of separating the variables was applied in order to solve it using three-dimensional spherical coordinates. It produces two equations, the first of which is a radial equation and the second of which is an angular equation. The radial equation's solution requires knowledge of the potential, while the angular equation's solution utilizes the special functions, notably the associated Legendre equation \cite{griffiths2018introduction}.\\
L'Hospital questioned in 1695 what it meant $\frac {d^n f}{dx^n}$ if $n=\frac{1}{2}$. A fractional derivative has since been described in some research studies. The majority of them defined the fractional derivative in integral form. For a fractional derivative, there are numerous definitions. Hadamard, Riemann-Liouville, Caputo, Riesz, Weyl, Grünwald, and Riesz-Caputo\cite{caputoLinearModelsDissipation1967,oldhamFractionalCalculusTheory1974,millerIntroductionFractionalIntegrals1993,kilbasTheoryApplicationsFractional,hilferApplicationsFractionalCalculus2000,podlubnyFractionalDifferentialEquations1998,klimekLagrangeanHamiltonianFractional2002,chenInvestigationFractionalFractal2010,heNewFractalDerivation2011,gorenfloEssentialsFractionalCalculus2000,kimeuFractionalCalculusDefinitions2009}.\\
The use of fractional calculus has recently become one of the most fascinating issues in a variety of physical science domains. In the last few decades, a hell of a lot of work on fractional calculus with various definitions has been produced; for example \cite{rabei_hamilton-jacobi_2012,rabei_hamiltonjacobi_2007,rabei_potentials_2004,baleanu_fractional_2008,rabei_hamilton_2007,rabei2009fractional}.\\
A novel derivative idea, the conformable derivative, was presented a few years ago by Khalil et al. \cite{khalil2014new}. The conformable  derivative of $f$ with order $0<\alpha \leq 1$ is defined by \cite{khalil2014new}
\be
\label{conformable}
T_\alpha(f)(t)=\lim_{\epsilon \to 0}\frac{f(t+\epsilon t^{1-\alpha})-f(t)}{\epsilon},
\ee
where $f\in [0,\infty) \to R$. This definition generally satisfies the standard properties of the traditional derivative, which makes it attractive for researchers. Some of these properties are \cite{khalil2014new}
\begin{itemize}
\item $D^{\alpha}(af+bg)=aT_\alpha(f)+bT_\alpha(g) $\quad for all real constant $a,b$ 
 
\item $D^{\alpha}(f g)=f T_\alpha(g)+g T_\alpha(f)$
 
\item $D^{\alpha}(t^p)=pt^{p-\alpha } $ for all $p$
 
 \item $D^{\alpha}(\frac{f}{g})=\frac{g T_\alpha(f)-f T_\alpha(g)}{g^2}$
 
 \item $D^{\alpha}(c)=0 $ with  $c$ is constant.
\end{itemize}
Also, one can easily verify  that
\begin{equation}
\label{D1}
D^{\alpha}[{\psi}(s)]=s^{1-\alpha}{\psi}(s)
\end{equation}
and
\begin{equation}
\label{D2}
D^{\alpha}[D^{\alpha}{\psi}(s)]=(1-\alpha)s^{1-2\alpha}\psi^{\prime}(s)+s^{2-2\alpha}\psi^{\prime\prime}(s).
\end{equation}\\
 For further knowledge about properties and applications of this type of derivative, we refer you to \cite{abdeljawad2015conformable,atangana2015new,khalil2014conformable,khalil2019geometric,zhao2017general,abu2019laguerre} and references therein. 
The conformable derivative does not satisfy zero-order, semigroup, or the Generalized Leibniz rule, but the conformable fractional derivative still contains components of the ordinary derivative. The conformable derivative is hence called a local operator \cite{teodoro2019review}.\\
The conformable calculus was in special relativity to study the effect of deformation of special relativity studied  by conformable derivative \cite{al-jamel_effect_2022}, and in quantum mechanics to study its effect on the formation of quantum-mechanical operators \cite{chung2020effect}, and the annihilation and creation operators are used to quantize  the conformable harmonic oscillator \cite{AlMasaeedRabeiAlJamelBaleanu+2021+395+401}. Recently it was used to extend the approximation methods in quantum mechanics (variational method \cite{al2022extension}, perturbation theory \cite{al2021extension} and WKB approximation \cite{al2021wkb}  ).\\
The purpose of this paper is to solve the conformable Schrodinger equation for the   hydrogen atom With conformable potential and energy levels solutions to study the behavior of its solutions with different  values of $\alpha$. 
%%%%%%%%%%%%%%%%%%%%%%%%%%%%%%%%%%%%%%%%%%%%%%%%%%%%%%%%%%%%%%%%%%%%%%%%%%%%%%%%%%%%%%%%%%%%%%%%%%%%%%%%%%%%%%%%%%%%%%%%%%%%%%%%%%%%%%%%%%%%%%%%%%
%%%%%%%%%%%%%%%%%%%%%%%%%%%%%%%%%%%%%%%%%%%%%%%%%%%%%%%%%%%%%%%%%%%%%%%%%%%%%%
%%%%%%%%%%%%%%%%%%%%%%%%%%%%%%%%%%%%%%%%%%%%%%%%%%%%%%%%%%%%%%%%%%%%%%%%%%%%%%%%
\section{Conformable Schrodinger Equation For Hydrogen atom}
The conformable Schrodinger equation in 3D-spherical coordinates is given by \cite{rabei2022solution}
\be
\label{shr 3d}
 \left( \nabla^{2\alpha} -\frac{2m^\alpha}{\hbar_\alpha^{2\alpha}}(V_\alpha(r^\alpha) - E^\alpha ) \right) \psi_\alpha(r^\alpha,\theta^\alpha,\varphi^\alpha)=0.
\ee
After using the separation of variables $\psi_\alpha(r^\alpha,\theta^\alpha,\varphi^\alpha) = R_{n \ell \alpha}(r^\alpha) Y_{\ell\alpha}^{m\alpha}$, we get two equations, the first one is reads as 
\be
\label{Y}
\frac{1}{Y_{\ell\alpha}^{m\alpha} \sin{(\theta^\alpha)}} D_\theta^\alpha [\sin{(\theta^\alpha)} D_\theta^\alpha Y_{\ell\alpha}^{m\alpha}] +\frac{1}{Y_{\ell\alpha}^{m\alpha} \sin^2{(\theta^\alpha)}} D_{\varphi}^{2\alpha} Y_{\ell\alpha}^{m\alpha}  = -\alpha^2 \ell(\ell+1).
\ee
this equation is called conformable angular equation  of the Schrodinger equation and its solution is given by \cite{rabei2022solution}
 \be
 \label{angular wf}
 Y_{\ell\alpha}^{m\alpha}=\sqrt{\frac{(2\ell+1)(\ell-m)!}{\alpha^{2m-2}2(\ell+m)!(2 \pi)^\alpha}} e^{i m \varphi^\alpha} P_{\ell \alpha}^{m \alpha}(\cos{(\theta^\alpha)}).
 \ee
The second equation is called  conformable radial equation is given as \cite{rabei2022solution}
\be
\label{radial}
  D_r^\alpha [r^{2\alpha} D_r^\alpha R_\alpha]+ \left[ \frac{2m^\alpha r^{2\alpha} }{\hbar_\alpha^{2\alpha}}(  E^\alpha - V_\alpha(r^\alpha)) - \alpha^2 \ell(\ell+1)\right]R_\alpha=0.
\ee
The solution for this equation is depends on knowing the conformable potential $V_\alpha(r^\alpha)$. So, for Hydrogen atom the conformable potential is reads as \cite{al2022extension}
\be
\label{vp}
V_\alpha(r^\alpha) = - \frac{a^\alpha}{r^\alpha},
\ee
where $a = \frac{e^2}{4 \pi \epsilon_0}$, after substituting in eq.\eqref{radial}, we get 
\be
\label{radial1}
  D_r^\alpha [r^{2\alpha} D_r^\alpha R_\alpha]+ \left[ \frac{2m^\alpha r^{2\alpha} }{\hbar_\alpha^{2\alpha}}(  E^\alpha + \frac{a^\alpha}{r^\alpha}) - \alpha^2 \ell(\ell+1)\right]R_\alpha=0.
\ee
let $R_\alpha(r^\alpha) = \frac{u_\alpha(r^\alpha)}{r^\alpha}$, we get 
\be
\label{radial2}
  D_r^\alpha D_r^\alpha u_\alpha + \left[ - k^2  + \frac{2m^\alpha  }{\hbar_\alpha^{2\alpha}} \frac{a^\alpha}{r^\alpha} -\frac{ \alpha^2 \ell(\ell+1)}{r^{2\alpha}} \right]u_\alpha=0.
\ee
where $k^2 = -\frac{2m^\alpha  E^\alpha }{\hbar_\alpha^{2\alpha}} $. Let $r^\alpha = \frac{\rho^\alpha }{2k}  \to D^\alpha_r = 2 k D^\alpha_\rho$, so, we get 
\be
\label{radial3}
 4 k^2 D^\alpha_\rho D^\alpha_\rho u_\alpha + \left[ - k^2  + \frac{2m^\alpha  }{\hbar_\alpha^{2\alpha}} \frac{a^\alpha 2k}{\rho^\alpha} -\frac{ 4 k^2 \alpha^2 \ell(\ell+1)}{\rho^{2\alpha}} \right]u_\alpha=0.
\ee
Thus, 
\be
\label{radial3}
  D^\alpha_\rho D^\alpha_\rho u_\alpha + \left[ - \frac{1}{4}  +  \frac{\lambda_\alpha }{ \rho^\alpha} -\frac{  \alpha^2 \ell(\ell+1)}{\rho^{2\alpha}} \right]u_\alpha=0.
\ee
where $\lambda_\alpha = \frac{m^\alpha a^\alpha  }{\hbar_\alpha^{2\alpha} k} $.
We solve this equation in two cases.
\\
\textbf{first case}: $\rho^\alpha \to 0 $, we get 
\be
\label{zero case}
  \rho^{2\alpha} D^\alpha_\rho D^\alpha_\rho u_\alpha -  \alpha^2 \ell(\ell+1) u_\alpha=0.
\ee
So, the solution for this equation is given by 
\be
\label{sol zero case}
u_\alpha(\rho^\alpha) = A \rho^{\alpha(\ell+1)}.
\ee

\textbf{Second case}: $\rho^\alpha \to \infty $, we get 
\be
\label{infty case}
  D^\alpha_\rho D^\alpha_\rho u_\alpha - \frac{1}{4} u_\alpha=0.
\ee
So, the solution for this equation is given by 
\be
\label{sol infty case}
u_\alpha(\rho^\alpha) = C \exp{\left(-\frac{\rho^\alpha}{2 \alpha} \right)}.
\ee
Thus, the approximate solution for  eq.\eqref{radial3} in two boundary conditions in eqs.\eqref{sol zero case} and \eqref{sol infty case} case . So,  we assume a general solution of the form $v_\alpha(\rho^\alpha) $
\be
\label{sol new}
u_\alpha(\rho^\alpha) = A \rho^{\alpha(\ell+1)} \exp{\left(-\frac{\rho^\alpha}{2 \alpha} \right)} v_\alpha(\rho^\alpha).
\ee
After substituting in eq.\eqref{radial3}, we get
\be
\label{v eq}
\rho^\alpha  D^\alpha_\rho D^\alpha_\rho v_\alpha + \left[ 2\alpha \ell + 2\alpha - \rho^\alpha \right] D^\alpha_\rho v_\alpha + \left[\lambda_\alpha -\alpha(\ell+1) \right]v_\alpha  = 0.
\ee
This equation corresponds to the Conformable Associated Laguerre equation. So the solution of this equation is given by Conformable Associated Laguerre functions \cite{rabei2021solution}
\be
\label{sol }
  L_{s \alpha}^{m}\left(\frac{ x^\alpha}{\alpha}\right)= \frac{x^{- m \alpha }\exp{\left(\frac{x^\alpha}{\alpha} \right)}}{\alpha^s  s!} D^{s\alpha} \left[x^{(s+m)\alpha}\exp{\left(-\frac{x^\alpha}{\alpha} \right)}\right].
\ee
where $m=2\ell+1, \quad \lambda_\alpha = n \alpha ,\quad s = n-\ell-1 $.\\
So, we can be written  the solution in eq.\eqref{sol new} as 
\be
\label{sol final}
u_\alpha(\rho^\alpha) = A \rho^{\alpha(\ell+1)} \exp{\left(-\frac{\rho^\alpha}{2 \alpha} \right)} L_{(n-\ell-1) \alpha}^{2\ell+1} \left(\frac{ \rho^\alpha}{\alpha}\right).
\ee
A can be  calculated constant $A$ using conformable normalization condition \cite{chung2020effect}, we get 
\bea
\nn
\int_0^\infty |u_\alpha(\rho^\alpha)|^2 \frac{d^\alpha \rho}{2 k}&=&\frac{|A|^2}{2k} \int_0^\infty   
 \rho^{2\alpha(\ell+1)} \exp{\left(-\frac{\rho^\alpha}{\alpha} \right)} \left[L_{(n-\ell-1) \alpha}^{2\ell+1} \left(\frac{ \rho^\alpha}{\alpha}\right)\right]^2 d^\alpha \rho \\\label{int1}&=& 1.
\eea
to solve this integral we Use the following relation which given by   \cite{rabei2021solution}
\bea
\label{some rel 8}
\int_0^\infty e^{-\frac{s^\alpha}{\alpha}}  x^{m\alpha+\alpha} L_{s \alpha}^m(\frac{ x^\alpha}{\alpha})  L_{k \alpha}^m(\frac{ x^\alpha}{\alpha}) d^\alpha x = \frac{\alpha^{m+1} (m+s)!}{s!} [2s+m+1].
\eea
So, the constant $A$ can be  calculated as  
\be
\label{nor cons}
A=\sqrt{\frac{k(n-\ell-1)!}{n \alpha^{2\ell+2}(n+\ell)!}}.
\ee
Thus, $\frac{k}{n}= \frac{1}{\alpha r_b^\alpha n^2}$, where  $r_b^\alpha$ called $\alpha-$ Bohr radius it is equal  $r_b^\alpha =\frac{(4\pi \epsilon_0)^\alpha \hbar_\alpha^{2\alpha} }{m^\alpha e^{2\alpha}}$. So, the eq.\eqref{sol final} becomes 
\be
\label{sol final 1}
u_\alpha(\rho^\alpha) =\sqrt{\frac{(n-\ell-1)!}{  r_b^\alpha n^2 \alpha^{2\ell+3}(n+\ell)!}} \rho^{\alpha(\ell+1)} \exp{\left(-\frac{\rho^\alpha}{2 \alpha} \right)} L_{(n-\ell-1) \alpha}^{2\ell+1} \left(\frac{ \rho^\alpha}{\alpha}\right).
\ee
As a result, the conformable radial wave function is equal 
\bea
\label{sol final 2}
R_\alpha(\rho^\alpha) &=&2 k  \frac{u_\alpha(\rho^\alpha)}{\rho^\alpha},\\\nn&=&\sqrt{\left(\frac{2}{\alpha n r_b^\alpha}\right)^3\frac{(n-\ell-1)!}{   2n \alpha^{2\ell+2}(n+\ell)!}} \rho^{\alpha \ell} \exp{\left(-\frac{\rho^\alpha}{2 \alpha} \right)} L_{(n-\ell-1) \alpha}^{2\ell+1} \left(\frac{ \rho^\alpha}{\alpha}\right)
\eea
After re-substituting  $ \rho^\alpha=\frac{ 2r^\alpha }{\alpha r_b^\alpha n}$, we obtain
\be
\label{sol final 2}
R_\alpha(r^\alpha) = \sqrt{\left(\frac{2}{\alpha n r_b^\alpha}\right)^3\frac{(n-\ell-1)!}{   2n \alpha^{2\ell+2}(n+\ell)!}} \left[\frac{ 2r^\alpha }{\alpha r_b^\alpha n}\right]^{\ell} \exp{\left(-\frac{ r^\alpha }{\alpha^2 r_b^\alpha n} \right)} L_{(n-\ell-1) \alpha}^{2\ell+1} \left(\frac{ 2r^\alpha }{\alpha^2 r_b^\alpha n}\right).
\ee
\begin{table}[htb!]
     \centering
      \caption{the first nine conformable spherical harmonics $R_\alpha(r^\alpha)$ }
     \begin{tabular}{c|c|c}
        n &  $\ell$ & $R_\alpha(r^\alpha)$ \\ \hline &&\\
         1 & 0& $\sqrt{\frac{4 }{\alpha^5 r_b^{3\alpha}}} \exp{\left(-\frac{ r^\alpha }{\alpha^2 r_b^\alpha } \right)} $\\ &&\\
           \multirow{2}{*}{2} & 0&$\sqrt{\frac{1}{8\alpha^5 r_b^{3\alpha}}}\left[2- \frac{ r^\alpha }{\alpha^2 r_b^\alpha }\right] \exp{\left(-\frac{ r^\alpha }{2\alpha^2 r_b^\alpha } \right)}$
           \\& 1& $\sqrt{\frac{1}{6\alpha^9 r_b^{5\alpha} }} \frac{r^\alpha}{2} \exp{\left(-\frac{ r^\alpha }{2\alpha^2 r_b^\alpha } \right)} $\\&&\\
         \multirow{3}{*}{3}&0&$\sqrt{\frac{4}{3^3\alpha^5 r_b^{3\alpha} }} \left[ \frac{2}{3}\left(\frac{ r^\alpha }{3\alpha^2 r_b^\alpha } \right)^2-\frac{ 2r^\alpha }{3\alpha^2 r_b^\alpha }  +1\right] \exp{\left(-\frac{ r^\alpha }{3\alpha^2 r_b^\alpha } \right)}$\\
         &1&$\sqrt{\frac{8}{3^7\alpha^9 r_b^{5\alpha} }} r^\alpha \left[ 2-\frac{ r^\alpha }{3\alpha^2 r_b^\alpha }
         \right]\exp{\left(-\frac{ r^\alpha }{3\alpha^2 r_b^\alpha } \right)}$
         \\&2& $\sqrt{\frac{1}{10 \alpha^9 3^5 r_b^{3\alpha} }}\left[\frac{ 2r^\alpha }{\alpha r_b^\alpha 3}\right]^{2} \exp{\left(-\frac{ r^\alpha }{3\alpha^2 r_b^\alpha } \right)}$
     \end{tabular}
     \label{tab:my_label}
 \end{table}
\newpage
\begin{figure}[htb!]
    \centering
    \includegraphics{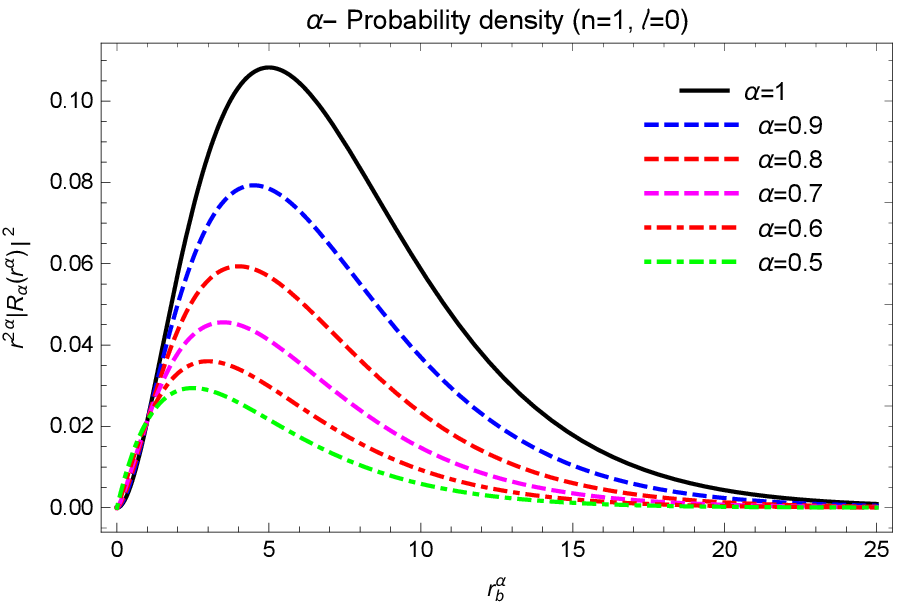}
    \caption{The $\alpha-$probability density $r^{2\alpha}|R_\alpha(r^\alpha)|^2$  at different values of $\alpha$. }
    \label{fig:my_label}
\end{figure}
\begin{figure}[htb!]
    \centering
    \includegraphics{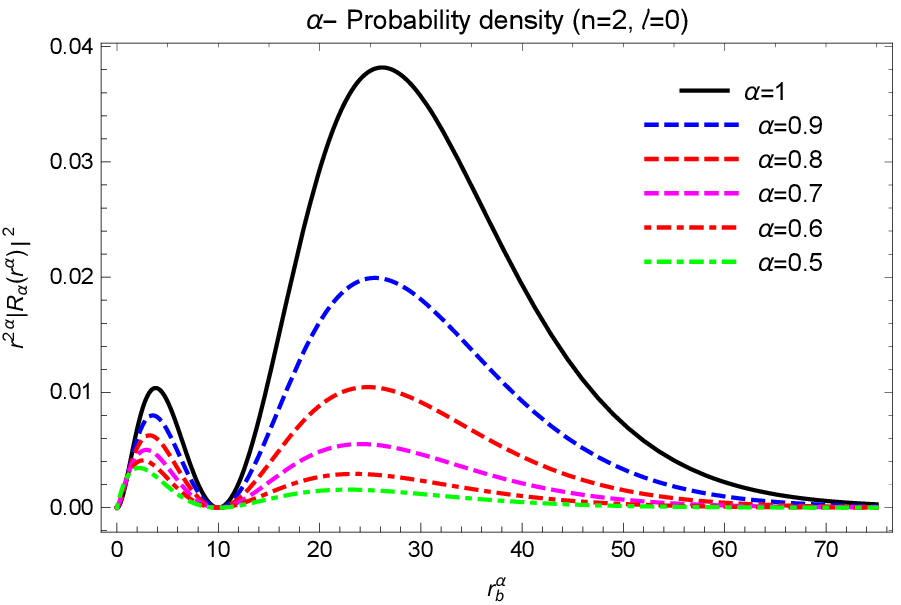}
    \caption{The $\alpha-$probability density $r^{2\alpha}|R_\alpha(r^\alpha)|^2$  at different values of $\alpha$. }
    \label{fig:my_label}
\end{figure}
\newpage
\begin{figure}[htb!]
    \centering
    \includegraphics{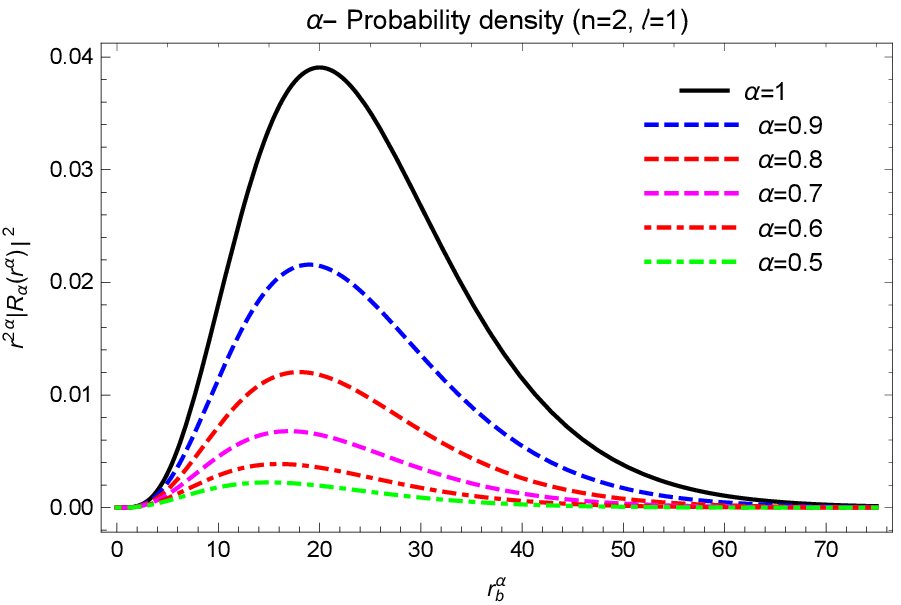}
    \caption{The $\alpha-$probability density $r^{2\alpha}|R_\alpha(r^\alpha)|^2$  at different values of $\alpha$. }
    \label{fig:my_label}
\end{figure}
\begin{figure}[htb!]
    \centering
    \includegraphics{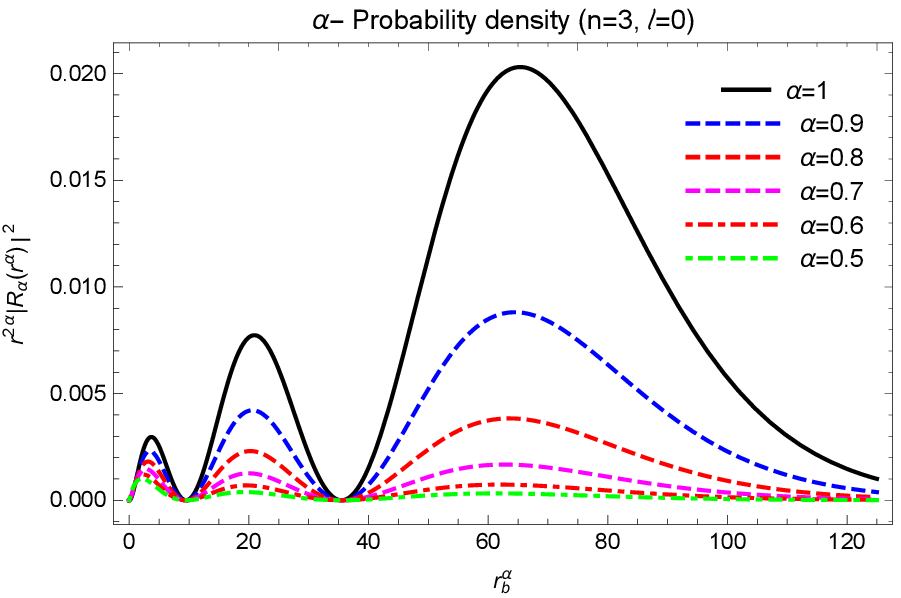}
    \caption{The $\alpha-$probability density $r^{2\alpha}|R_\alpha(r^\alpha)|^2$  at different values of $\alpha$. }
    \label{fig:my_label}
\end{figure}
\newpage
\begin{figure}[htb!]
    \centering
    \includegraphics{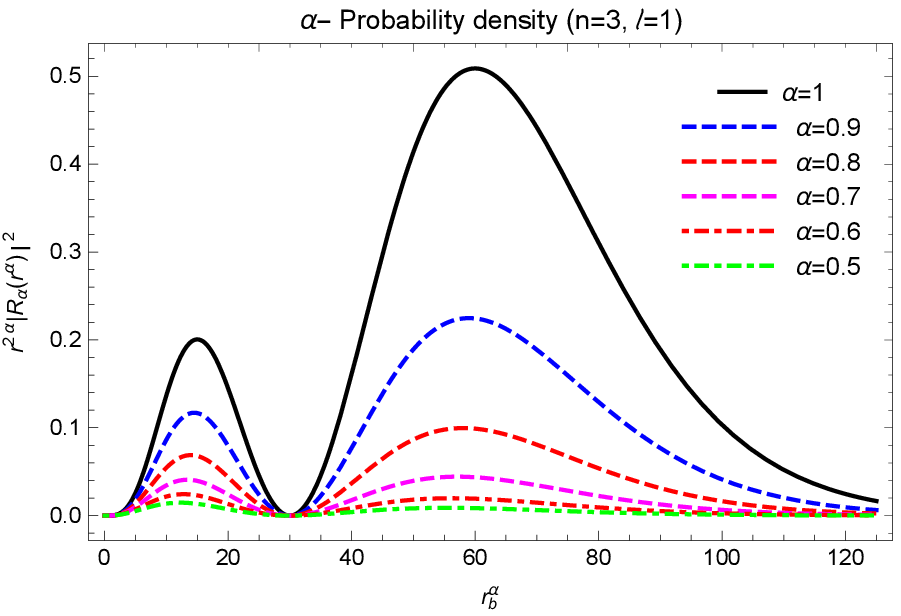}
    \caption{The $\alpha-$probability density $r^{2\alpha}|R_\alpha(r^\alpha)|^2$  at different values of $\alpha$. }
    \label{fig:my_label}
\end{figure}
\begin{figure}[htb!]
    \centering
    \includegraphics{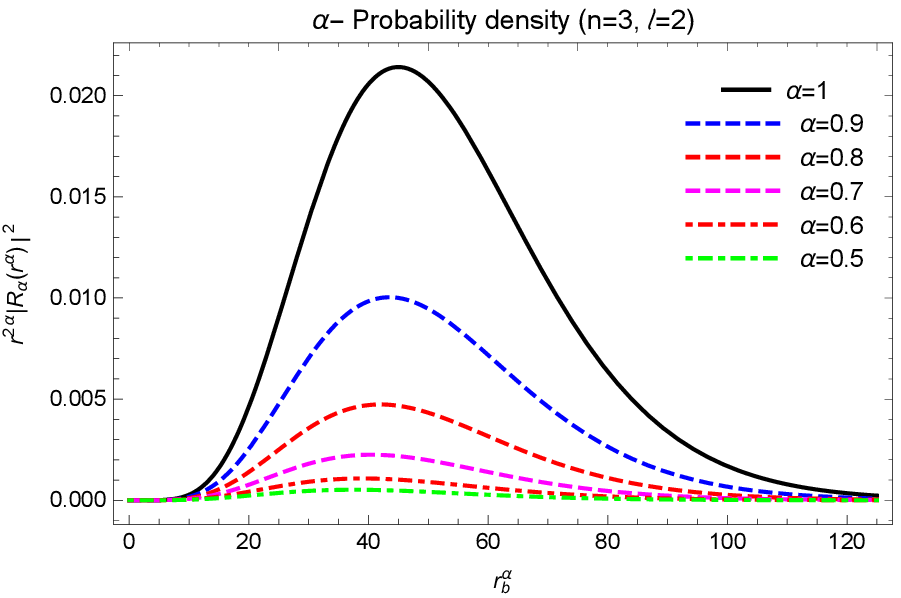}
    \caption{The $\alpha-$probability density $r^{2\alpha}|R_\alpha(r^\alpha)|^2$  at different values of $\alpha$. }
    \label{fig:my_label}
\end{figure}
Thus, the $\alpha-$energy levels are taken from this formula 
\bea
E^\alpha  =-  \frac{(13.6 ev)^\alpha}{2^{1-\alpha}\alpha^2  n^2} 
\eea
where $\frac{m^\alpha}{2^\alpha \hbar_\alpha^{2\alpha} } \left( \frac{e^{2}}{4\pi \epsilon_0}\right)^{2\alpha}= (13.6 ev)^\alpha $, 
\begin{figure}[htb!]
    \centering
    \includegraphics{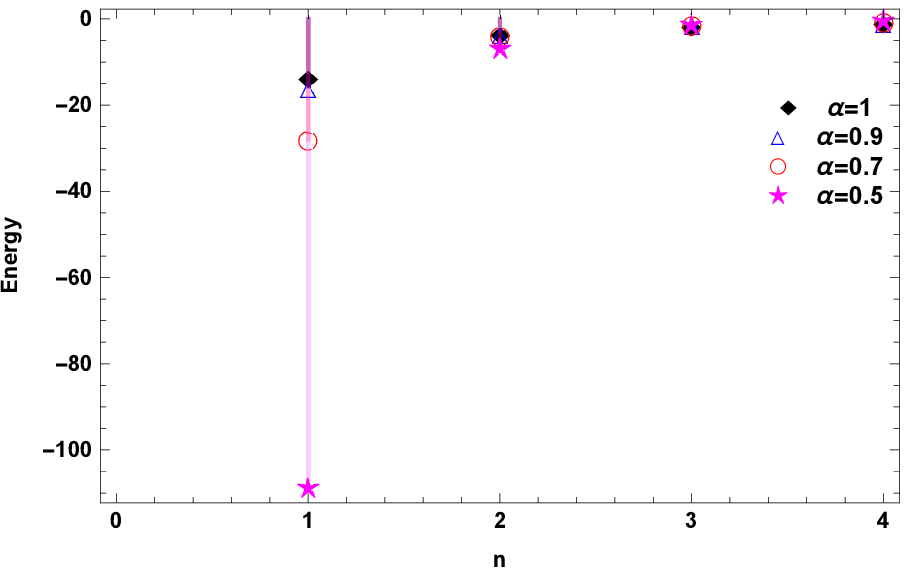}
    \caption{The $\alpha-$energy levels $E^\alpha$ as a function of quantum number n  at different values of $\alpha$. }
    \label{fig:my_label}
\end{figure}
\begin{figure}[htb!]
    \centering
    \includegraphics{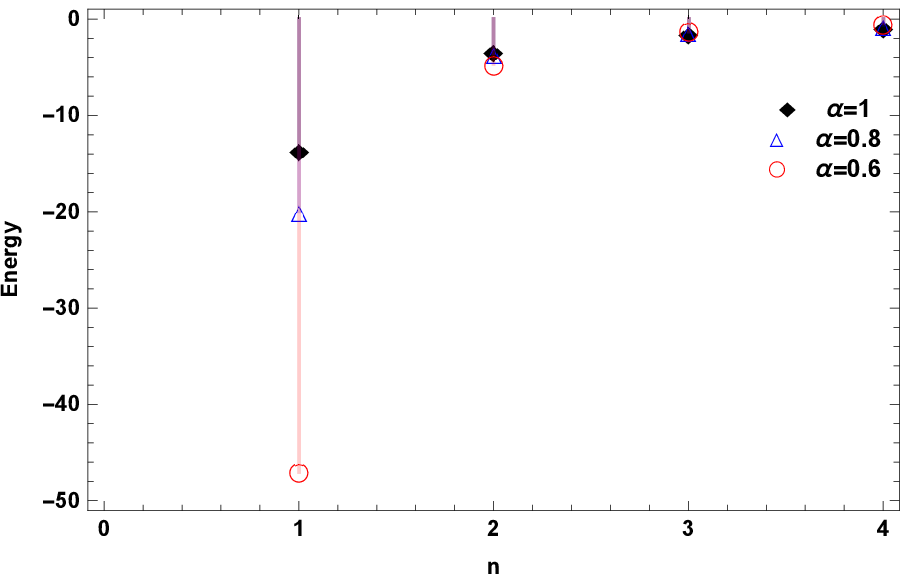}
    \caption{The $\alpha-$energy levels $E^\alpha$ as a function of quantum number n  at different values of $\alpha$. }
    \label{fig:my_label}
\end{figure}\newpage
So, the solution for the conformable Schrodinger equation for the Hydrogen atom is given by
\bea
\psi_{n \ell m \alpha}(r^\alpha,\theta^\alpha,\varphi^\alpha) &=& R_{n \ell \alpha}(r^\alpha) Y_{\ell\alpha}^{m\alpha}\\\nn
&=&\sqrt{\left(\frac{1}{\alpha n r_b^\alpha}\right)^3\frac{2(n-\ell-1)!(2\ell+1)(\ell-m)!}{   n \alpha^{2\ell+2m}(n+\ell)!(\ell+m)!(2 \pi)^\alpha}} \left[\frac{ 2r^\alpha }{\alpha r_b^\alpha n}\right]^{\ell} \exp{\left(-\frac{ r^\alpha }{\alpha^2 r_b^\alpha n} \right)}\\\nn&& L_{(n-\ell-1) \alpha}^{2\ell+1} \left(\frac{ 2r^\alpha }{\alpha^2 r_b^\alpha n}\right) e^{i m \varphi^\alpha} P_{\ell \alpha}^{m \alpha}(\cos{(\theta^\alpha)})
\eea
\newpage
\begin{table}[htb!]
     \centering
      \caption{the  conformable wave function $\psi_{n \ell m \alpha}(r^\alpha,\theta^\alpha,\varphi^\alpha)$ for different values }
     \begin{tabular}{c|c|c|c}
         n&$\ell$ & $m$ & $\psi_{n \ell m \alpha}(r^\alpha,\theta^\alpha,\varphi^\alpha) $ \\ \hline &&\\
        1& 0 & 0& $\sqrt{\frac{2 }{\alpha^3 r_b^{3\alpha }(2 \pi)^\alpha}} \exp{\left(-\frac{ r^\alpha }{\alpha^2 r_b^\alpha } \right)}$\\ &&\\
        2& 0&0& $\sqrt{\frac{1}{16(2 \pi)^\alpha\alpha^3 r_b^{3\alpha}}}\left[2- \frac{ r^\alpha }{\alpha^2 r_b^\alpha }\right] \exp{\left(-\frac{ r^\alpha }{2\alpha^2 r_b^\alpha } \right)}$\\
        &1 &0& $  \sqrt{\frac{1}{4\alpha^7 r_b^{5\alpha} (2 \pi)^\alpha}} \frac{r^\alpha}{2} \exp{\left(-\frac{ r^\alpha }{2\alpha^2 r_b^\alpha } \right)}\cos{(\theta^\alpha)} $\\
         &1&1& $-  \sqrt{\frac{1}{8\alpha^7 r_b^{5\alpha} (2 \pi)^\alpha}} \frac{r^\alpha}{2} \exp{\left(-\frac{ r^\alpha }{2\alpha^2 r_b^\alpha } \right)} e^{i\varphi^\alpha}\sin{(\theta^\alpha)}$
     \end{tabular}
     \label{tab:my_label}
 \end{table}

\begin{figure}[htb!]
    \centering
    \includegraphics{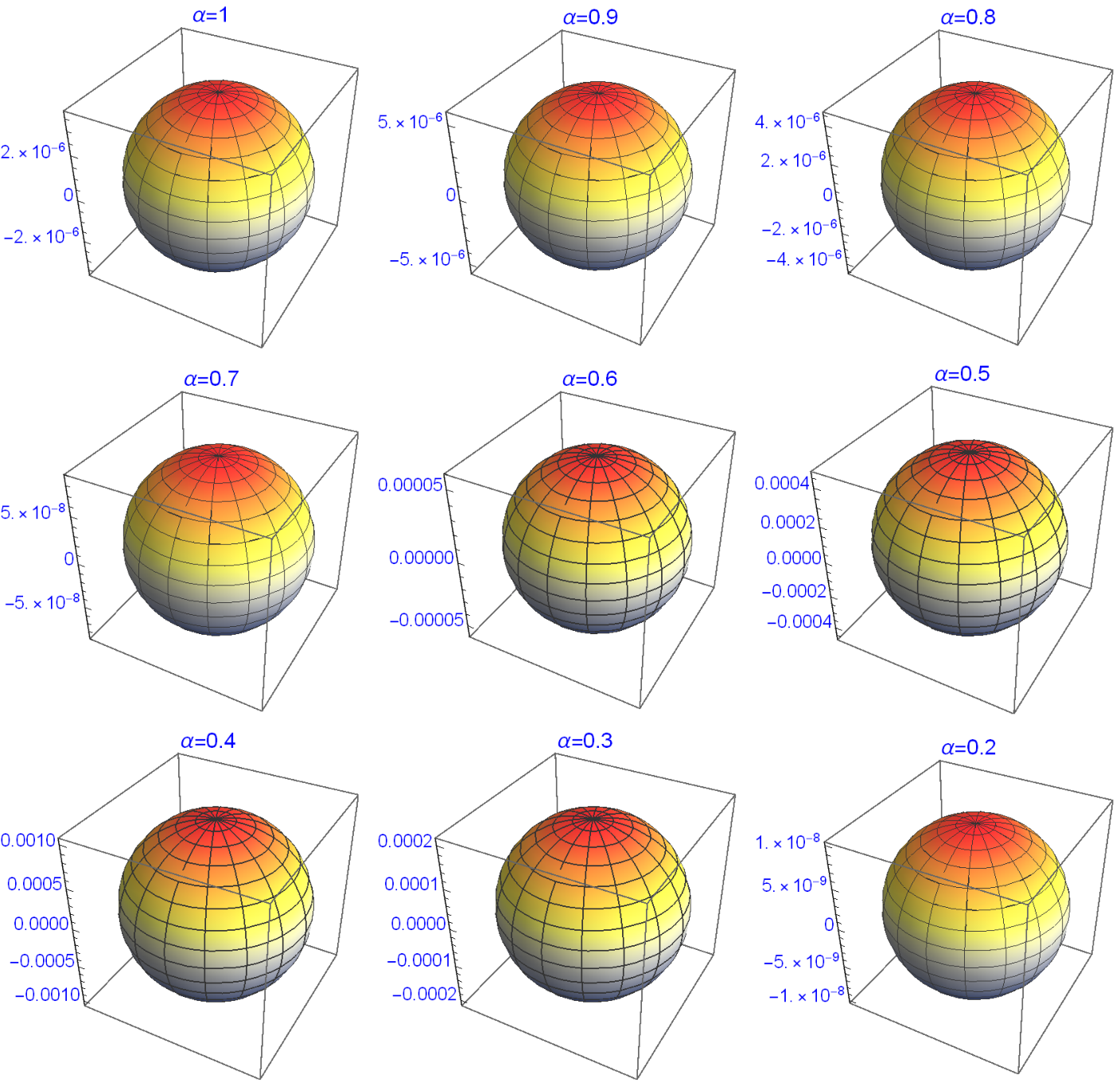}
    \caption{ Plot $|\psi_{200 \alpha}|^2$ with different value of $\alpha$ from 0.1 to 0.2 in 3d}
    \label{fig:my_label}
\end{figure}
\begin{figure}[htb!]
    \centering
    \includegraphics{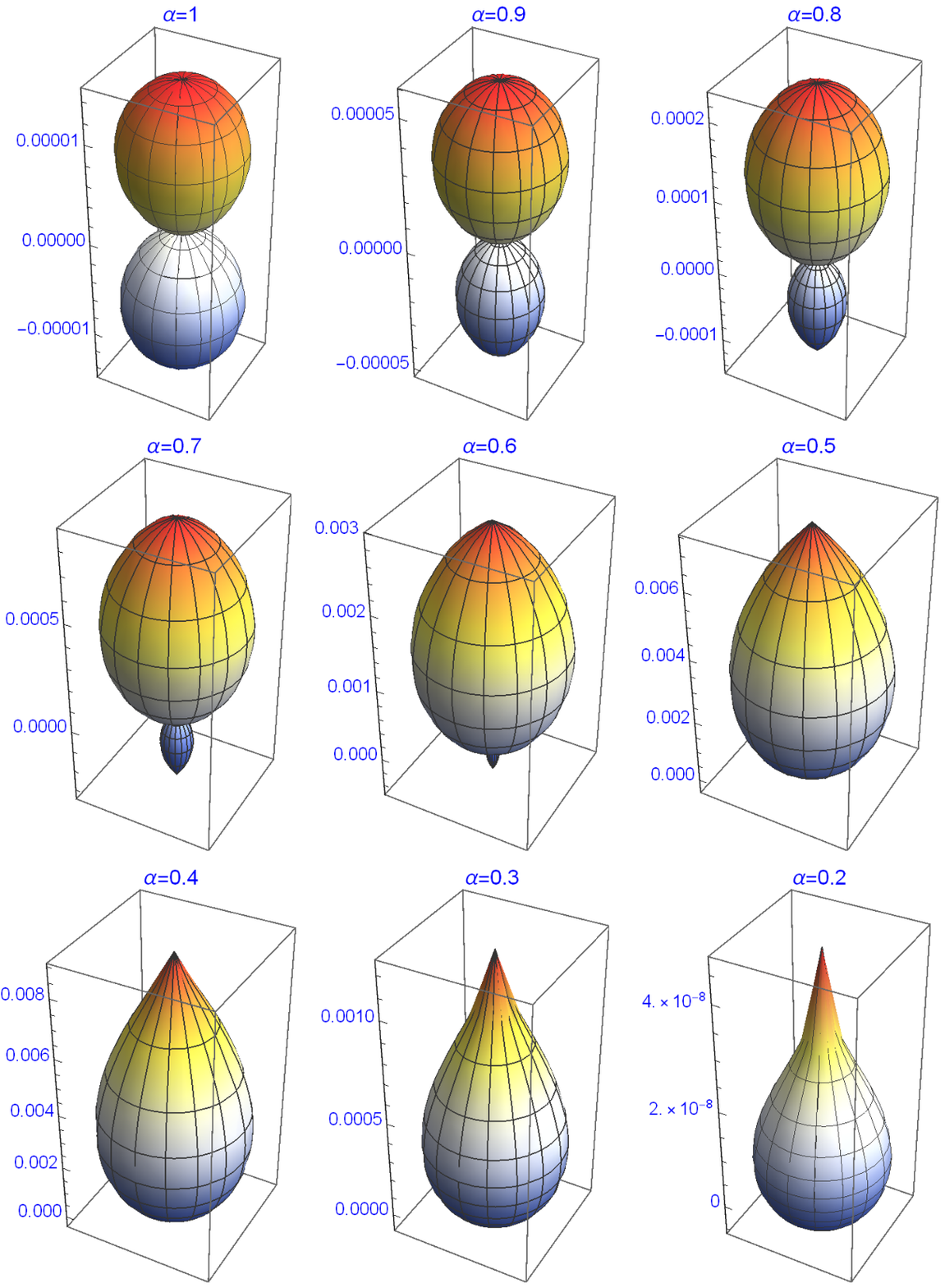}
    \caption{ Plot $|\psi_{210 \alpha}|^2$ with different value of $\alpha$ from 0.1 to 0.2 in 3d}
    \label{fig:my_label}
\end{figure}
\begin{figure}[htb!]
    \centering
    \includegraphics{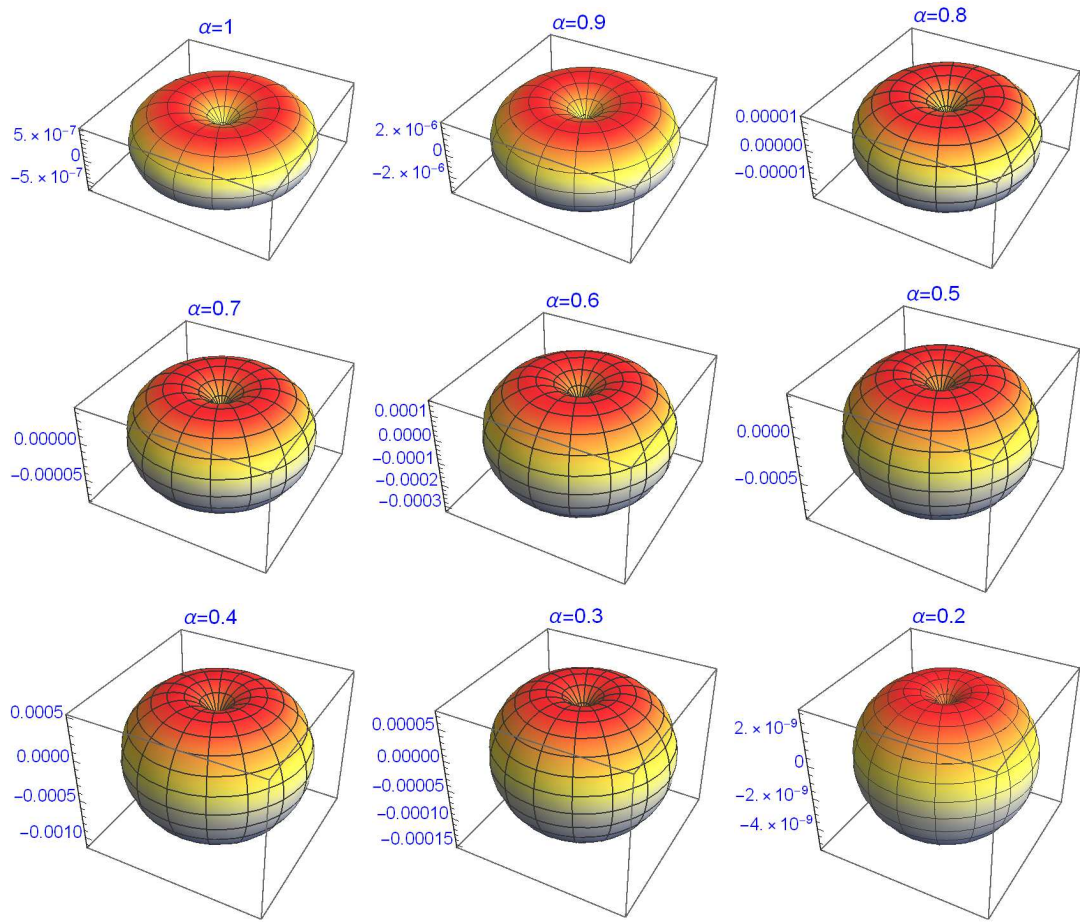}
    \caption{ Plot $|\psi_{211 \alpha}|^2$ with different value of $\alpha$ from 0.1 to 0.2 in 3d}
    \label{fig:my_label}
\end{figure}
\clearpage
%%%%%%%%%%%%%%%%%%%%%%%%%%%%%%%%%%%%%%%%%%%%%%%%%%%%%%%%%%%%%%%%%%%%%%%%%%%%%%%%%%%%%%%%%%%%%%%%%%%%%%%%%%%%%%%%%%%%%%%%%%%%%%%%%%%%%%%%%%%%%%\
\section{Conclusions}
The conformable Schrodinger equation in 3D-spherical coordinates is solved and wave functions and energy levels for different values of $\alpha$ are obtained. The conformable wave function for $n=1$ and  $n=2$ are calculated. it is observed that the traditional  wave function can be recovered  when $\alpha=1$.\\
In addition, the $\alpha$- probability density for $n=1, \ell=0$, $n=2,\ell=1$ and $n=3, \ell=0,1,2$ are drown  for different values of $\alpha$. it is concluded that the $\alpha$- probability density gradually converts to the traditional case.
%%%%%%%%%%%%%%%%%%%%%%%%%%%%%%%%%%%%%%%%%%%%%%%%%%%%%%%%%%%%%%%%%%%%%%%%%%%%%%
%%%%%%%%%%%%%%%%%%%%%%%%%%%%%%%%%%%%%%%%%%%%%%%%%%%%%%%%%%%%%%%%%%%%%%%%%%%%%%%%%%%%%%%%%%%%%%%%%%%%%%%%%%%%%%%%%%%%%%%%%%%%%%%%%%%%%%%%%%%%%%%%%%
\bibliography{Href}

\begin{thebibliography}{10}
\providecommand{\url}[1]{#1}
\csname url@samestyle\endcsname
\providecommand{\newblock}{\relax}
\providecommand{\bibinfo}[2]{#2}
\providecommand{\BIBentrySTDinterwordspacing}{\spaceskip=0pt\relax}
\providecommand{\BIBentryALTinterwordstretchfactor}{4}
\providecommand{\BIBentryALTinterwordspacing}{\spaceskip=\fontdimen2\font plus
\BIBentryALTinterwordstretchfactor\fontdimen3\font minus
  \fontdimen4\font\relax}
\providecommand{\BIBforeignlanguage}[2]{{%
\expandafter\ifx\csname l@#1\endcsname\relax
\typeout{** WARNING: IEEEtran.bst: No hyphenation pattern has been}%
\typeout{** loaded for the language `#1'. Using the pattern for}%
\typeout{** the default language instead.}%
\else
\language=\csname l@#1\endcsname
\fi
#2}}
\providecommand{\BIBdecl}{\relax}
\BIBdecl

\bibitem{griffiths2018introduction}
D.~J. Griffiths and D.~F. Schroeter, \emph{Introduction to quantum
  mechanics}.\hskip 1em plus 0.5em minus 0.4em\relax Cambridge University
  Press, 2018.

\bibitem{caputoLinearModelsDissipation1967}
M.~Caputo, ``Linear models of dissipation whose {Q} is almost frequency
  independent—{II},'' \emph{Geophysical Journal International}, vol.~13,
  no.~5, pp. 529--539, 1967, publisher: Blackwell Publishing Ltd Oxford, UK.

\bibitem{oldhamFractionalCalculusTheory1974}
K.~Oldham and J.~Spanier, \emph{The fractional calculus theory and applications
  of differentiation and integration to arbitrary order}.\hskip 1em plus 0.5em
  minus 0.4em\relax Elsevier, 1974.

\bibitem{millerIntroductionFractionalIntegrals1993}
K.~S. Miller and B.~Ross, \emph{An introduction to the fractional integrals and
  derivatives-theory and applications}.\hskip 1em plus 0.5em minus 0.4em\relax
  Wiley, New York, 1993.

\bibitem{kilbasTheoryApplicationsFractional}
A.~Kilbas, \emph{Theory and applications of fractional differential equations}.

\bibitem{hilferApplicationsFractionalCalculus2000}
R.~Hilfer, \emph{Applications of fractional calculus in physics}.\hskip 1em
  plus 0.5em minus 0.4em\relax World scientific Singapore, 2000, vol.~35,
  no.~12.

\bibitem{podlubnyFractionalDifferentialEquations1998}
I.~Podlubny, \emph{Fractional differential equations: an introduction to
  fractional derivatives, fractional differential equations, to methods of
  their solution and some of their applications}.\hskip 1em plus 0.5em minus
  0.4em\relax Elsevier, 1998.

\bibitem{klimekLagrangeanHamiltonianFractional2002}
M.~Klimek, ``Lagrangean and {Hamiltonian} fractional sequential mechanics,''
  \emph{Czechoslovak Journal of Physics}, vol.~52, no.~11, pp. 1247--1253,
  2002, publisher: Springer.

\bibitem{chenInvestigationFractionalFractal2010}
W.~Chen, X.-D. Zhang, and D.~Korošak, ``Investigation on fractional and
  fractal derivative relaxation-oscillation models,'' \emph{International
  Journal of Nonlinear Sciences and Numerical Simulation}, vol.~11, no.~1, pp.
  3--10, 2010, publisher: De Gruyter.

\bibitem{heNewFractalDerivation2011}
J.-H. He, ``A new fractal derivation,'' \emph{Thermal Science}, vol.~15, no.
  suppl. 1, pp. 145--147, 2011.

\bibitem{gorenfloEssentialsFractionalCalculus2000}
R.~Gorenflo and F.~Mainardi, ``Essentials of fractional calculus,'' 2000,
  publisher: Citeseer.

\bibitem{kimeuFractionalCalculusDefinitions2009}
J.~M. Kimeu, ``Fractional calculus: {Definitions} and applications,'' 2009.

\bibitem{rabei_hamilton-jacobi_2012}
\BIBentryALTinterwordspacing
E.~M. Rabei and B.~S. Ababneh, ``\BIBforeignlanguage{en}{Hamilton-{Jacobi}
  {Fractional} {Sequential} {Mechanics}},'' Jun. 2012, accepted:
  2012-06-08T14:16:15Z. [Online]. Available:
  \url{https://repositorio.leon.uia.mx/xmlui/handle/20.500.12152/40437}
\BIBentrySTDinterwordspacing

\bibitem{rabei_hamiltonjacobi_2007}
\BIBentryALTinterwordspacing
E.~M. Rabei, I.~Almayteh, S.~I. Muslih, and D.~Baleanu,
  ``\BIBforeignlanguage{en}{Hamilton–{Jacobi} formulation of systems within
  {Caputo}'s fractional derivative},'' \emph{\BIBforeignlanguage{en}{Physica
  Scripta}}, vol.~77, no.~1, p. 015101, 2007, publisher: IOP Publishing.
  [Online]. Available: \url{https://doi.org/10.1088/0031-8949/77/01/015101}
\BIBentrySTDinterwordspacing

\bibitem{rabei_potentials_2004}
\BIBentryALTinterwordspacing
E.~M. Rabei, T.~S. Alhalholy, and A.~Rousan, ``Potentials of arbitrary forces
  with fractional derivatives,'' \emph{International Journal of Modern Physics
  A}, vol.~19, no. 17n18, pp. 3083--3092, 2004, publisher: World Scientific
  Publishing Co. [Online]. Available:
  \url{https://www.worldscientific.com/doi/abs/10.1142/S0217751X04019408}
\BIBentrySTDinterwordspacing

\bibitem{baleanu_fractional_2008}
\BIBentryALTinterwordspacing
D.~Baleanu, S.~I. Muslih, and E.~M. Rabei, ``\BIBforeignlanguage{en}{On
  fractional {Euler}–{Lagrange} and {Hamilton} equations and the fractional
  generalization of total time derivative},''
  \emph{\BIBforeignlanguage{en}{Nonlinear Dynamics}}, vol.~53, no.~1, pp.
  67--74, 2008. [Online]. Available:
  \url{https://doi.org/10.1007/s11071-007-9296-0}
\BIBentrySTDinterwordspacing

\bibitem{rabei_hamilton_2007}
\BIBentryALTinterwordspacing
E.~M. Rabei, K.~I. Nawafleh, R.~S. Hijjawi, S.~I. Muslih, and D.~Baleanu,
  ``\BIBforeignlanguage{en}{The {Hamilton} formalism with fractional
  derivatives},'' \emph{\BIBforeignlanguage{en}{Journal of Mathematical
  Analysis and Applications}}, vol. 327, no.~2, pp. 891--897, 2007. [Online].
  Available:
  \url{https://www.sciencedirect.com/science/article/pii/S0022247X06004525}
\BIBentrySTDinterwordspacing

\bibitem{rabei2009fractional}
E.~M. Rabei, I.~M. Altarazi, S.~I. Muslih, and D.~Baleanu, ``Fractional wkb
  approximation,'' \emph{Nonlinear Dynamics}, vol.~57, no. 1-2, pp. 171--175,
  2009.

\bibitem{khalil2014new}
R.~Khalil, M.~Al~Horani, A.~Yousef, and M.~Sababheh, ``A new definition of
  fractional derivative,'' \emph{Journal of Computational and Applied
  Mathematics}, vol. 264, pp. 65--70, 2014.

\bibitem{abdeljawad2015conformable}
T.~Abdeljawad, ``On conformable fractional calculus,'' \emph{Journal of
  computational and Applied Mathematics}, vol. 279, pp. 57--66, 2015.

\bibitem{atangana2015new}
A.~Atangana, D.~Baleanu, and A.~Alsaedi, ``New properties of conformable
  derivative,'' \emph{Open Mathematics}, vol.~1, no. open-issue, 2015.

\bibitem{khalil2014conformable}
R.~Khalil and M.~Abu-Hammad, ``Conformable fractional heat differential
  equation,'' \emph{International Journal of Pure and Applied Mathematics},
  vol.~94, pp. 215--217, 2014.

\bibitem{khalil2019geometric}
R.~Khalil and M.~A. H. M.~A. Hammad, ``Geometric meaning of conformable
  derivative via fractional cords,'' \emph{J. Math. Computer Sci}, vol.~19, pp.
  241--245, 2019.

\bibitem{zhao2017general}
D.~Zhao and M.~Luo, ``General conformable fractional derivative and its
  physical interpretation,'' \emph{Calcolo}, vol.~54, no.~3, pp. 903--917,
  2017.

\bibitem{abu2019laguerre}
M.~Abu~Hammad, B.~Albarmawi, A.~Shmasneh, A.~Dababneh \emph{et~al.}, ``Laguerre
  equation and fractional laguerre polynomials,'' \emph{J. Semigroup Theory
  Appl.}, vol. 2019, pp. Article--ID, 2019.

\bibitem{teodoro2019review}
G.~S. Teodoro, J.~T. Machado, and E.~C. De~Oliveira, ``A review of definitions
  of fractional derivatives and other operators,'' \emph{Journal of
  Computational Physics}, vol. 388, pp. 195--208, 2019.

\bibitem{al-jamel_effect_2022}
\BIBentryALTinterwordspacing
A.~Al-Jamel, M.~Al-Masaeed, E.~Rabei, and D.~Baleanu,
  ``\BIBforeignlanguage{en}{The effect of deformation of special relativity by
  conformable derivative},'' \emph{\BIBforeignlanguage{en}{Revista Mexicana de
  Física}}, vol.~68, no. 5 Sep-Oct, pp. 050\,705 1--9, Aug. 2022, number: 5
  Sep-Oct. [Online]. Available:
  \url{https://rmf.smf.mx/ojs/index.php/rmf/article/view/5877}
\BIBentrySTDinterwordspacing

\bibitem{chung2020effect}
W.~S. Chung, S.~Zare, H.~Hassanabadi, and E.~Maghsoodi, ``The effect of
  fractional calculus on the formation of quantum-mechanical operators,''
  \emph{Mathematical Methods in the Applied Sciences}, 2020.

\bibitem{AlMasaeedRabeiAlJamelBaleanu+2021+395+401}
\BIBentryALTinterwordspacing
M.~Al-Masaeed, E.~M. Rabei, A.~Al-Jamel, and D.~Baleanu, ``Quantization of
  fractional harmonic oscillator using creation and annihilation operators,''
  \emph{Open Physics}, vol.~19, no.~1, pp. 395--401, 2021. [Online]. Available:
  \url{https://doi.org/10.1515/phys-2021-0035}
\BIBentrySTDinterwordspacing

\bibitem{al2022extension}
M.~Al-Masaeed, E.~M. Rabei, and A.~Al-Jamel, ``Extension of the variational
  method to conformable quantum mechanics,'' \emph{Mathematical Methods in the
  Applied Sciences}, vol.~45, no.~5, pp. 2910--2920, 2022.

\bibitem{al2021extension}
M.~Al-Masaeed, E.~M. Rabei, A.~Al-Jamel, and D.~Baleanu, ``Extension of
  perturbation theory to quantum systems with conformable derivative,''
  \emph{Modern Physics Letters A}, p. 2150228, 2021.

\bibitem{al2021wkb}
M.~Al-Masaeed, E.~Rabei, A.~Al-Jamel \emph{et~al.}, ``Wkb approximation with
  conformable operator,'' \emph{arXiv preprint arXiv:2111.01547}, 2021.

\bibitem{rabei2022solution}
E.~Rabei, M.~Al-Masaeed, A.~Al-Jamel \emph{et~al.}, ``Solution of the
  conformable angular equation of the schrodinger equation,'' \emph{arXiv
  preprint arXiv:2203.11615}, 2022.

\bibitem{rabei2021solution}
E.~Rabei, A.~Al-Jamel, M.~Al-Masaeed \emph{et~al.}, ``The solution of
  conformable laguerre differential equation using conformable laplace
  transform,'' \emph{arXiv preprint arXiv:2112.01322}, 2021.

\end{thebibliography}
\bibliographystyle{IEEEtran}
\end{document}